# Condensation of nucleoid in *Escherichia coli* cells as a result of prolonged starvation.

N.G. Loiko (1 and 2), Ya.A. Danilova (3), A.V. Moiseenko (3), E.V. Demkina (2), Kovalenko V. V(1)., K.B. Tereshkina (1), G.I. El-Registan (2), O.S. Sokolova (3) Yu.F. Krupyanskii (1)

((1) Semenov Institute of Chemical Physics RAS, (2) Federal Research Center "Fundamentals of Biotechnology" RAS, (3) Lomonosov Moscow State University)

Electron microscopy and X-ray diffraction studies of dormant *E. coli* cells revealed several forms of nucleoid condensation: quasi - nanocrystalline, quasi - liquid crystalline (or spore-like) and folded nucleosome - like structure. Of particular interest is the third type of structure since it was described here for the first time: the folded nucleosome-like. Such a structure has no relation to the toroidal DNA organization, which are the intermediate form in the formation of the quasi - nanocrystalline structure. Results observed here shed a new light both on the phenomenon of nucleoid condensation in prokaryotic cells and on the general problem of developing a response to stress. It was found out that there is no single mechanism for nucleoid condensation in the population of a dormant cell; diversity in their number, shape and packing has been seen. According to the recognized concept of a bacterial population as a multicellular organism, its heterogeneity allows to respond flexibly to the environmental changes and to survive in stressful situations. That is the reason why we observed at least three types of nucleoid condensation in dormant *E. coli* cells. Heterogeneity of dormant cells increases the ability of the whole population to survive under various stress conditions. For better understanding of the nucleoid condensation mechanism, it is necessary to study the reverse transition of the nucleoid in bacteria from the dormant to the functional state.

Introduction

Adaptive molecular strategies that ensure the ability of microorganisms to survive in environments significantly different from those optimal for their growth have been the subject of active research over the past years (Frenkiel-Krispin et al., 2001; Krupyanskii et al., 2018; Minsky et al., 2002; Sinitsyn et al., 2017; Tkachenko, 2018; Wolf et al., 1999; Zorraquino et al., 2017). Among them, the universal adaptive response of microorganisms to starvation is of specific interest, because it often coincides with the antibiotics resistance of pathogenic bacteria, which represents one of the most important medical problems in the world. Adaptive strategy may launch the increasing synthesis (up to 150-200 thousand copies) of a ferritin-like protein, Dps, are often involved in stress response (Chiancone & Ceci, 2010; De Martino et al., 2016; Talukder & Ishihama, 2014; Talukder et al., 1999). Increased synthesis in cells of E. coli occurs in the late stationary phase, under starving conditions.

Dps is a DNA-binding protein that carries on a regulatory and protective role in *E.coli* cells (Almirón et al., 1992). Its structure (Grant et al., 1998) and interactions with DNA were recently excessively studied *in vivo, in vitro* (Calhoun & Kwon, 2011; Frenkiel-Krispin & Minsky, 2006), and *in silico* (Antipov et al., 2017; Tereshkin et al., 2018). It has been noticed that the expression of Dps in bacterial cells improves their survival not only during starvation, but also protects them from oxidative stress, heat, acid, alkaline shock, toxic effects of heavy metals, antibiotics, UV radiation, etc. (Calhoun & Kwon, 2011; Karas, Westerlaken, & Meyer, 2015; Nair & Finkel, 2004; Zeth, 2012).

The generalization and analysis of a vast array of data on the biology of bacteria has led recently to the emergence of the concept that considers the microbial population as a multicellular organism (Bacteria as multicellular organism, 1997). According to this concept, the properties of cells in a population are heterogeneous; therefore, one may expect the different structural adaptive response to stress for the different cell of a population.

One of the main stress-protective function of Dps is to provide the condensation of DNA into "biocrystalline" or *in cellulo* nanocrystalline structure (Wolf et al., 1999, Frenkiel-Krispin et al., 2001; Frenkiel-Krispin et al., 2004; Doye & Poon, 2006; Karas et al., 2015; Krupyanskii et al., 2018; Minsky et al., 2002; Lee et al., 2015;Sinitsyn et al., 2017; Tkachenko, 2018; Krupyanskii et al., 2018; Sinitsyn et al., 2017). The structures formed by nucleoid in *E.coli* cells with overexpression of Dps within the starvation period up to 48 hours, were investigated by a number of methods (Frenkiel-Krispin et al., 2004; Krupyanskii et al., 2018; Minsky et al., 2002; Sinitsyn et al., 2017; Loiko et al., 2017). After 24 hours of starvation, toroidal structures of DNA were observed in the cytoplasm (Frenkiel-Krispin et al., 2004). Then (48 hours of starvation) nearly periodic structure appeared. Such a "biocrystals" were stable against the influence of nucleases and oxidants as well (Minsky et al., 2002; Wolf et al., 1999).

It was observed also that in starved bacteria that lack the *dps* gene, the DNA is reorganized from the apparently random distribution that is seen in actively growing cells into a cholesteric liquid-crystalline structure (Minsky et al., 2002;). The tight DNA packaging in this condensed phase reduces the accessibility of DNA molecules to various damaging factors, as well as in the case for DNA–Dps co-crystals.

The aim of the present study is to define the possible structures of *E. coli* cells for different strains, different Dps expression vectors, and different growing conditions under the stress of prolonged starvation.

## Results.

Preparation of samles.

To identify the nucleoid structures in the cytoplasm of *E.coli* cells, we compared the following strains: K12; Top10, Top10/pBAD-DPS and BL21-Gold(DE3)/pET-DPS (see the Table). The two latter strains were cultivated with and without induction of the Dps protein overproduction.

| Sample description | | | | Types of condensed structure, % of cells in the population | | |
|---|---|---|---|---|---|---|
| E.coli strain | Cultivation conditions | | | # quasi – nanocrystals | # quasi - liquid crystals | # folded nucleosome-like |
| | Growth medium | Age, months | Dps overexpression | | | |
| K12 | M9 | 7 | - | 50 | 35 | 15 |
| Top 10 | LB | 7 | - | 73 | 22 | 5 |
| Top10/pBAD-DPS | LB | 7 | - | 78 | 17 | 5 |
| | M9 | 7 | - | 29 | 55 | 16 |
| | LB | 7 | + | 85 | 9 | 6 |
| | M9 | 7 | + | 36 | 41 | 23 |
| BL21-Gold(DE3)/pET-DPS | M9 | 7 | + | 55 | 27 | 18 |

Table.

Types of condensed nucleoid structures in studied *E.coli* dormant cells

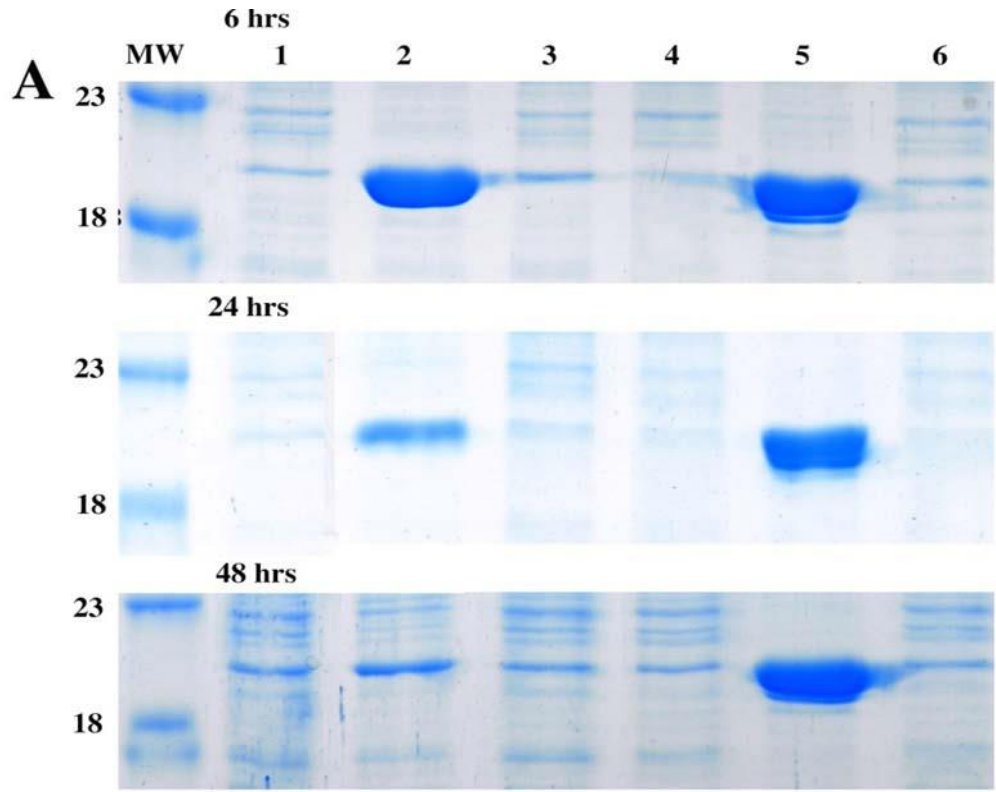

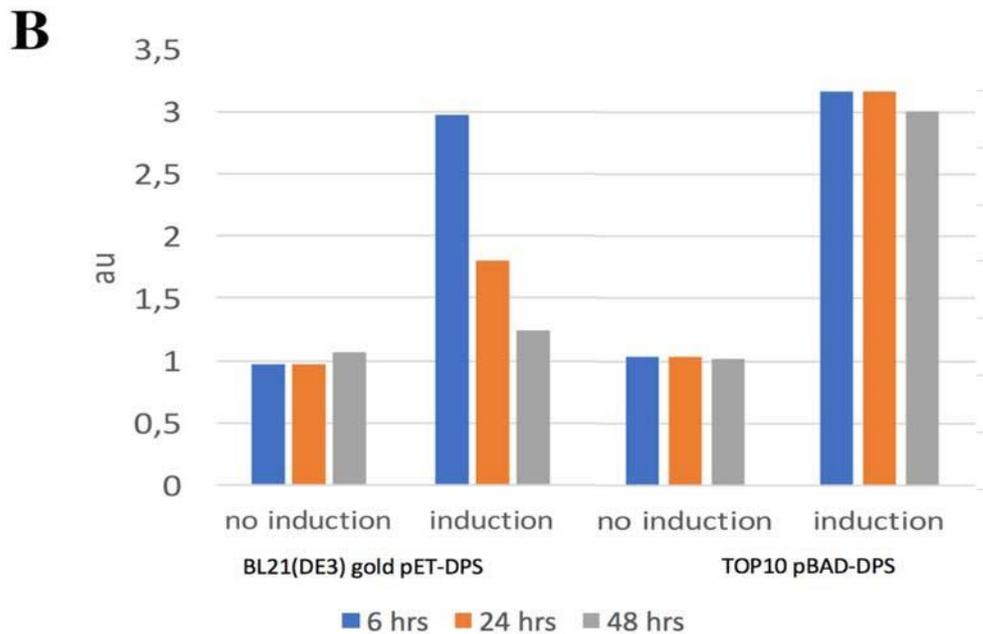

Fig.1 The levels of Dps expression for 6, 24 and 48 hrs without induction and after induction. Results of electrophoresis in SDS-PAAG for two strains BL21- Gold (DE3) /pET-DPS and Top10/pBAD-DPS.

The overproduction of Dps was induced by arabinose in the Top10/pBAD-DPS strain, and lactose in the BL21- Gold (DE3) / pET-DPS strain. Both inductions have been accomplished during the linear growth phase after 24 hrs of cultivation. The levels of Dps expression for 6-48 hrs after induction have been tested using the electrophoresis in SDS-PAAG (Fig. 1A). Both strains do not show a pronounced expression of Dps before induction (Fig. 1B), but six hours after induction the Dps expression level increased ~ 3 times in both strains. Top10/pBAD-DPS strain has the maximum expression of Dps at 24 hrs after induction. To obtain the dormant cells, both strains were removed from the shaker and stored for 7 months at 21ºC. All dormant cells were characterized by a lack of metabolism, a resistance to stress, an altered structure and the ability to grow in reversion. These resemble all properties of anabiotic forms of gram-negative microorganisms (Suzina et al., 2006).

We were able to observe three types of nucleoid structure in dormant *E. coli* cells by analyzing the ultrathin sections of fixed and Epon 812-embedded samples and X-ray diffraction: quasi - nanocrystalline, quasi -liquid crystalline and folded nucleosome - like structure (Fig. 2 - 5). The percentage of specific structure type varied significantly depending on the strain and cultivation conditions (Table). Thus, in the populations of the anabiotic cells of strains Top10 and Top10/pBAD-DPS without overexpression of Dps, that were cultivated on LB medium, the cells with quasi- nanocrystalline structure were predominant (more than 70%). Induction of Dps biosynthesis led to the increase of this cells up to ~85%. On the contrary, the dormant forms of Top10/pBAD-DPS strains obtained on synthetic M9 medium have only 40% cells bearing quasi- nanocrystalline structure, around 40% of them have quasi-liquid crystalline structure. Maximum number of cells bearing folded nucleosome - like structure were detected in the strain Top10/pBAD-DPS with overexpression of Dps, growing on synthetic M9 medium.

Quasi - nanocrystalline is the most frequent structure found earlier in bacterial cells starving for up to 48 h (Frenkiel-Krispin & Minsky, 2006; Minsky et al.,

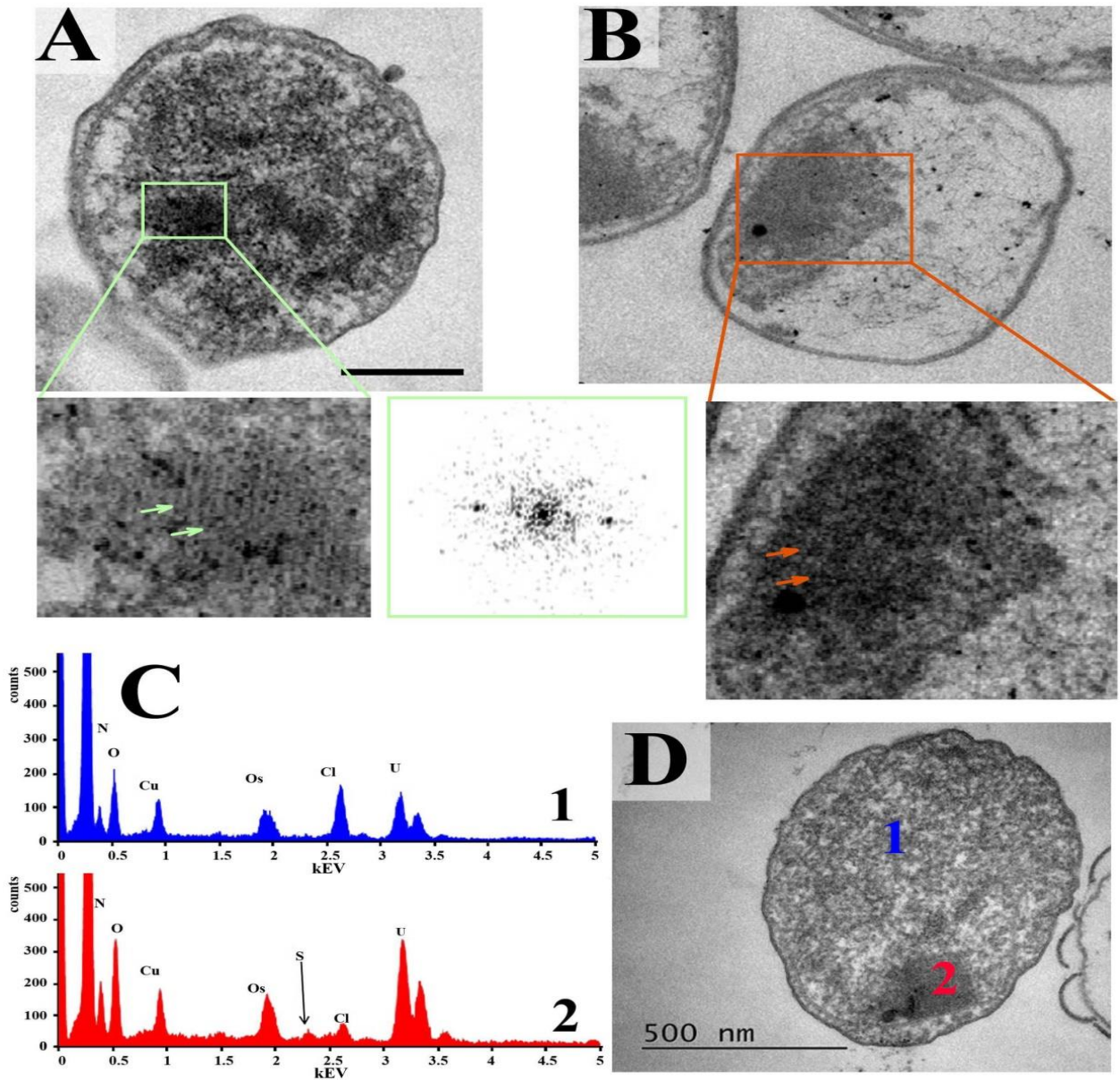

Fig. 2. Morphology of Dps–DNA quasi - crystals in dormant *E. coli* cells. (A) *E.coli* BL21-Gold(DE3)/pET-DPS growing on M9 media with induced Dps production in the linear growth phase, age 7 months; insert – assembly, close to crystalline, right – FFT from the selected in (A) area. (B) *E.coli* Top10/pBAD-DPS growing on LB media without inducing Dps production, age 7 months, insert – nearly amorphous assembly; arrows are pointing to the individual Dps–DNA layers. Bar – 250 nm. (C) Energy dispersive X-ray (EDX) spectra from the selected areas marked in (D) as: 1 - cytoplasm, and 2 – assembly, close to crystalline.

2002; Sinitsyn et al., 2017). Complex structure of a nucleoid which contains high polymer as DNA inside may consist of various regions, each with a characteristic degree of internal order (Kasai and Kakudo, 2005) , ranging continuously from some of it close to the ideal crystals to the completely amorphous state. Broad diffraction peaks (shown on Fig.3) are indicative of the imperfection (quasi-crystallinity) of condensed nucleoid. That is the reason why this state here will be called as quasi – nanocrystalline. These structures are thought to incorporate the bacterial nucleoid interconnected by a large number of

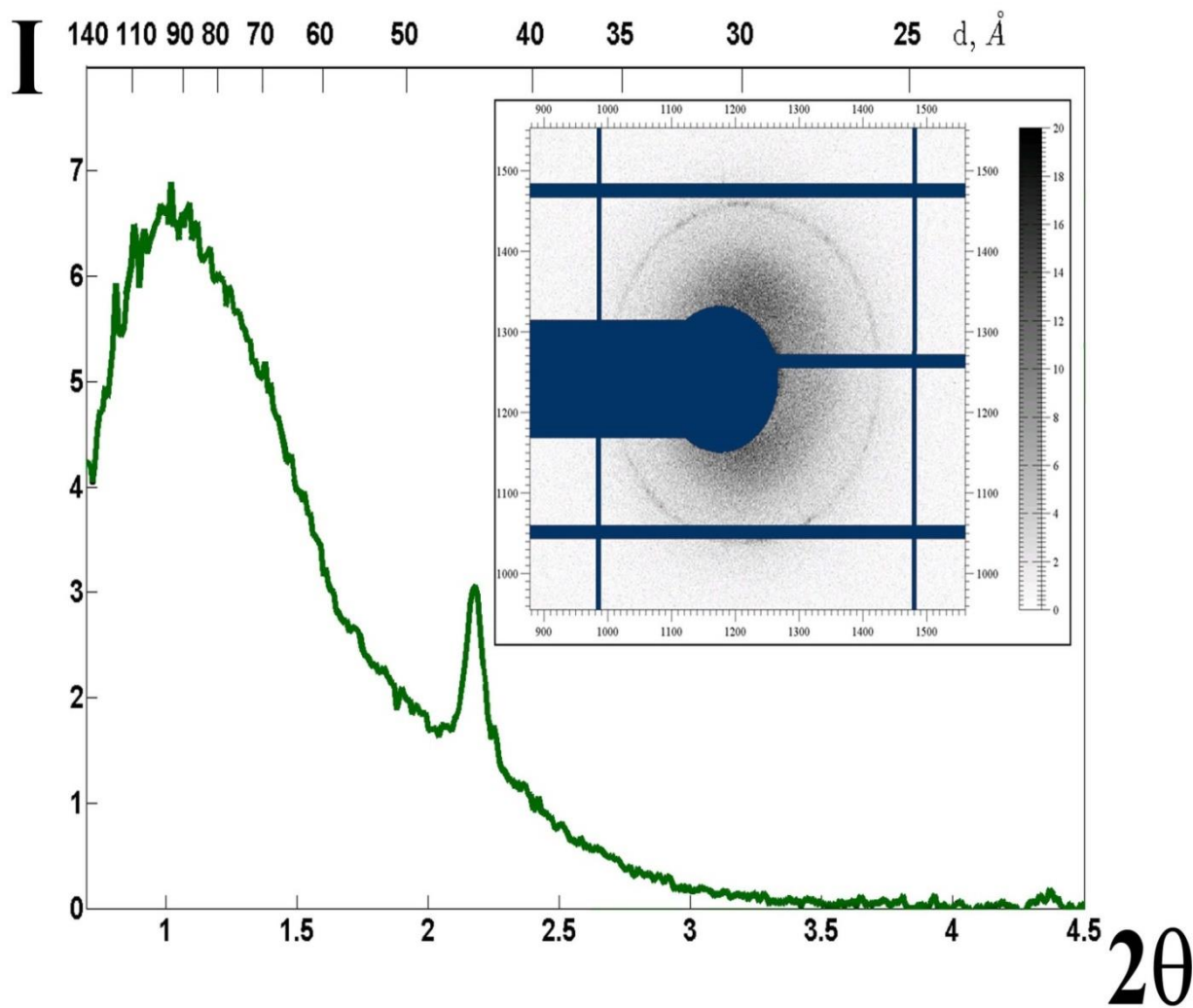

Fig.3. The dependence of the scattering intensity on the angle of 2Θ for the sample of starving bacteria E. coli (strain BL21- Gold (DE3) / pET-DPS). The insert shows the diffraction pattern for this sample. Complex structure of a nucleoid which contains high polymer as DNA inside may consist of various regions, each with a characteristic degree of internal order, ranging continuously from some of it close to the ideal crystals to the completely amorphous state. That is the reason why this state here will be called as quasi – nanocrystalline. Broad diffraction peaks (shown on Fig.3) are indicative of the imperfection (quasi-crystallinity) of condensed nucleoid.

Dps dodecamers. It was noticed that the size and number of quasi - nanocrystalline structures in dormant cells varied depending on the *E.coli* strain and Dps production. Thus, in *E.coli* strain BL21-Gold (DE3)/pET-DPS, grown with Dps induction on M9 media and aged 7 months, the number of quasi - nanocrystals usually varied from 5 to 10 per cell; with sizes approximately 40-80 nm (Fig. 2A). In the Top10/pBAD-DPS cells, grown without Dps induction on LB medium and aged 7 months, one quasi-crystal (size around 300-400 nm) occupied from one-quarter to one-half of one cell (Fig. 2B). Notably, upon overproduction of the Dps protein, the structure of condensed nucleoid became more ordered, closer to crystalline. Fourier transforms of such types of structure revealed not ideal crystalline lattice (Fig. 2A, insert; Fig. 4C, D). In the cells without overexpression of the Dps (Fig. 2B), the structure of condensed nucleoid is less ordered, not allowing to obtain the Fourier transform from these samples. The interlayer distance in this less ordered structures vary from 7 to 10 nm. It may be affected by a different number of Dps molecules that are interacting with DNA (Frenkiel-Krispin et al., 2004).

Analytical Electron Microscopy (EM) (precisely, energy dispersive X-ray spectroscopy (EDX)) was used to study the elemental composition of quasi - nanocrystalline structure (Fig. 2C). Here it was assumed that, within the bounding area, it is possible to detect the presence of sulfur, which, as suggested, reflects the existence of DPS (each dodecamer of Dps contains 48 Methionines), and phosphorus, which corresponds to DNA. Indeed, we detected the presence of sulfur in the areas of the cell that correspond to quasi - nanocrystalline formations (region 2 in Fig. 2D). As the control, we subjected the region outside the quasi – crystalline structure (region 1 in Fig. 2D) to elemental analysis and did not detect sulfur in this area, indicating that there is no increased concentration of Dps protein in the cytoplasm except of quasi - nanocrystalline region, which agrees with previous studies. Unexpectedly, we could not detect phosphorus in quasi – nanocrystalline region, apparently, its concentration was too low on the ultrathin section.

To obtain the interlayer distance of the quasi – nanocrystalline structure, double-axis tomography (Fig. 6C) was used. This made it possible to obtain an improved

Fourier transform and, by inverse Fourier transform, an image of a DNA-Dps nanocrystal filtered from noise (Fig. 5D). The distance between the crystal layers was around 7-8 nm, which is comparable but slightly less than in results obtained from cryo-EM (Frenkiel-Krispin et al., 2004), and less than characteristic DPS-DPS distance from the Fig.3 (first broad peak, corresponding to 9 nm), which may be the results of chemical fixation used.

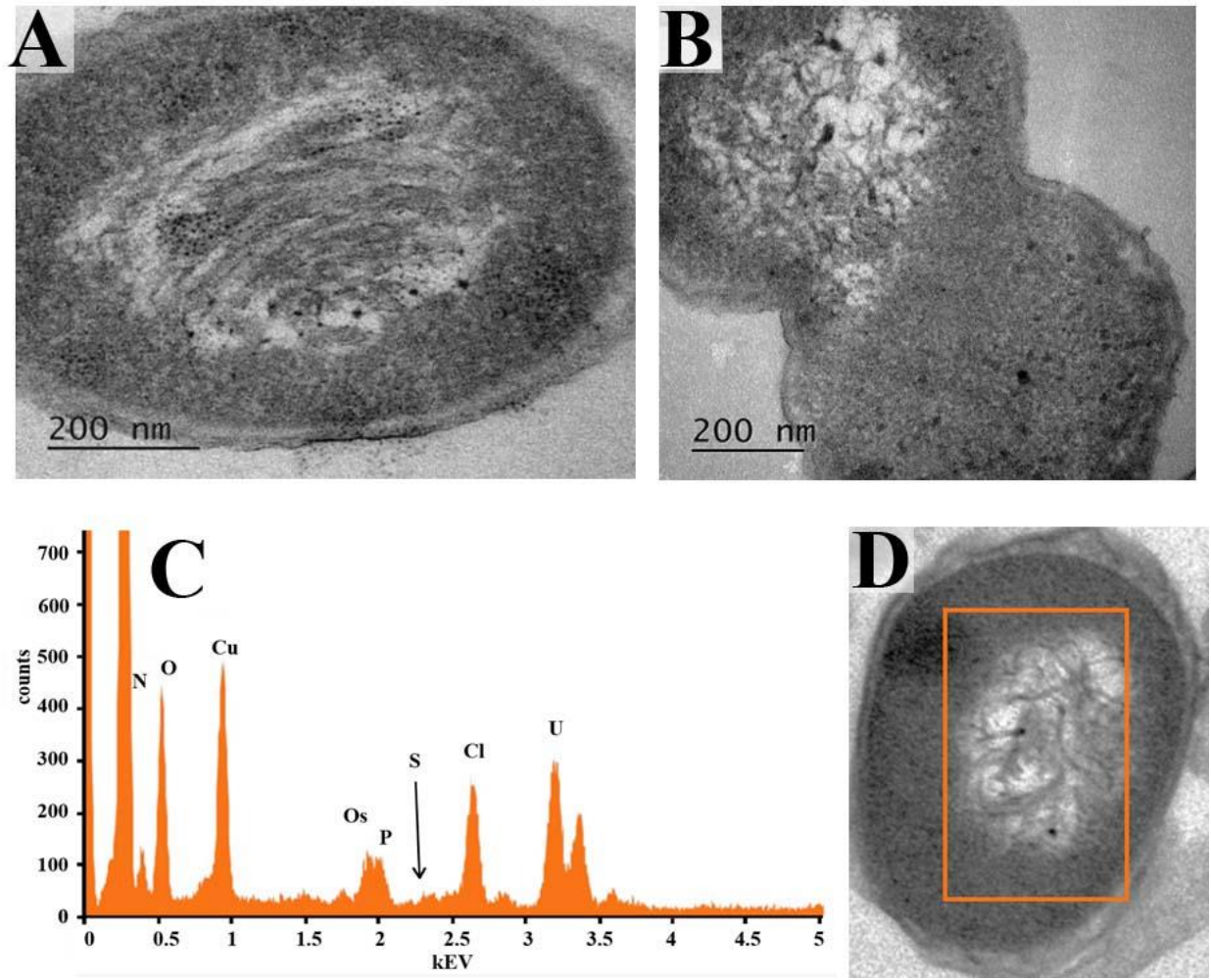

Fig. 4. Morphology of Dps–DNA quasi - liquid crystalline (or spore - like) assembly in dormant *E.coli* cells. (A) *E.coli* Top10/pBAD-DPS growing on M9 media, with induced Dps production in the linear growth phase, age 7 months; (B) *E. coli* BL21-Gold(DE3)/pET-DPS growing on M9 media, with induced Dps production in the linear growth phase, age 7 months; (C) EDX spectrum from the selected area, marked in (D) by orange rectangle. The positive detection of phosphorus, in this case, is due to a weak bound between the DNA and the protein. It should be noted that in the areas surrounding the liquid crystal, neither sulfur nor phosphorus have been identified.

Quasi - liquid crystalline (or spore-like) structures is the second recognized type of nucleoid condensed structure. This type of structure has been detected previously in starved *E.coli* lacking the *dps* gene(Minsky et al., 2002), as well as in spores of various origin (Loiko et al., 2017). Remarkably, each of the studied in present work *E.coli* populations possessed the *dps* gene. The number of cells bearing quasi - liquid crystalline structures in populations of dormant *E.coli* cells ranged from ~10% up to ~55% (Table) with the majority observed in the cells growing on M9 synthetic medium. In quasi- liquid crystalline structures, the nucleoid is placed in a central part of the cell, surrounded by condensed cytoplasm. In some cell the nucleoid is presented in the form close to parallel rod-like species, such as DNA molecules that are partially aligned in successive layers that continuously rotate with respect to each other. Such a DNA condensation in starved Dps– E. coli cells close to a cholesteric liquid-crystalline DNA phase (Fig. 4 A). The tight DNA packaging in this condensed phase reduces the accessibility of DNA molecules to various damaging factors, including irradiation, oxidizing agents and nucleases (Minsky et al., 2002). Another picture is for the cells displayed on the Fig. 4 (B, D). Such type of condenced DNA structures much closer to the spore-like structures (Loiko et al., 2017). In starved bacteria that have small amount of DPS, DNA molecules are randomly distributed. They are significantly less ordered than in cholesteric liquid - crystalline DNA phase (Fig. 4 A) and in previous quasi – nanocrystalline structure (Fig.2A). Small amount of DPS in liquid - crystalline DNA phase should results in low protein-to-DNA ratio in the chromatin of these cells.

The presence of Dps and DNA in liquid crystalline structure was confirmed by the elemental analysis of the central portion of the cells. Analysis shows the presence of both phosphorus and sulfur (Fig. 4 C, D). Amount of phosphorus is much larger than amount of sulfur (Fig. 4 C). The positive detection of phosphorus is due to a fact that the bound between the DNA and protein mainly weak amount of tightly bound to DNA protein is small. It should be noted that in the areas surrounding the

quasi - liquid crystalline structure, neither sulfur nor phosphorus have been identified.

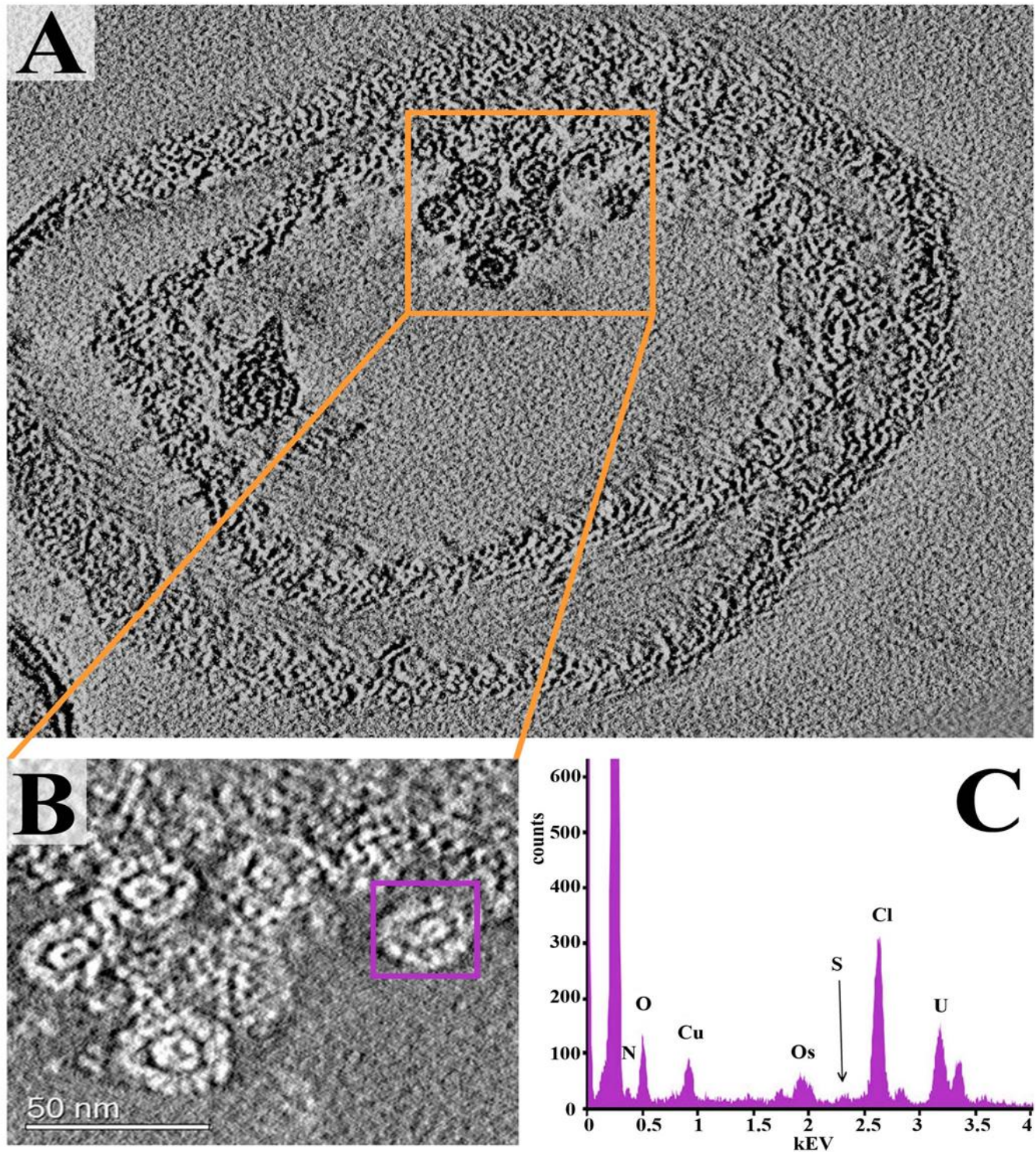

Fig. 5. Morphology of Dps - DNA folded nucleosome - like structure in dormant *E.coli* cells. (A)Tomography of *E.coli* cell strain Top10/pBAD-DPS, growing on M9 media, without induction of Dps production, age 7 months. Bar size – 500 nm; (B) Part selected in (A) with a bar size – 50 nm. (D) EDX spectrum from the area marked by white rectangle in (B). These *structures* are not the toroids described previously (Frenkiel-Krispin et al., 2004), but represent much smaller, nearly spherical formations.

Folded nucleosome - like structure has not previously been described in literature. In all studied populations of dormant bacteria cells, both with and without (Fig.5A) overproduction of Dps, 5% to 25% of cells (Table) contain round structures, with an average diameter 30 nm. The tomographic study (Fig. 5A,B) clearly demonstrated that these structures are not toroidal described previously (Frenkiel-Krispin et al., 2004), but represent much smaller and nearly spherical formations, This type of structure we have decided to call as folded nucleosome-like structure. It was suggested that in bacteria cells spherical aggregates of Dps molecules either may act similarly to histones, on which the DNA is twisted, or, which more probable, arrange the DPS beads (Fig.7) through which the DNA (string) passes (beads on the string Fig.8A). To counteract external stress factors, the beads should be placed quite tightly on the DNA (string). In addition, like in eukaryotic cells where the nucleosomes are folded to form fi ber loops, which then form the chromatid of a chromosome, beads on the string may folded into a globule-like structure. Possible schematic presentation of folded nucleosome - like structure, suggested here may be seen on Fig 8 (B, C,D). External adherent DPS molecules can additionally protect the DNA (Fig 8 C). Element analysis demonstrated that the spherical aggregates, indeed, contained sulfur (Fig. 5C), indicating the increased concentration of Dps protein. Unfortunately, as in the case of quasi-nanocrystalline structure, we failed to identify phosphorus. This is because of this condensed form where the DNA is tightly packed with the protein to counteract external stress factors.

It is necessary to note that in many cells, various types of above mentioned structures co-detected one with another, , suggesting that specific types of structure may be formed to protect different parts of the DNA, similar to eukaryotic chromatin.

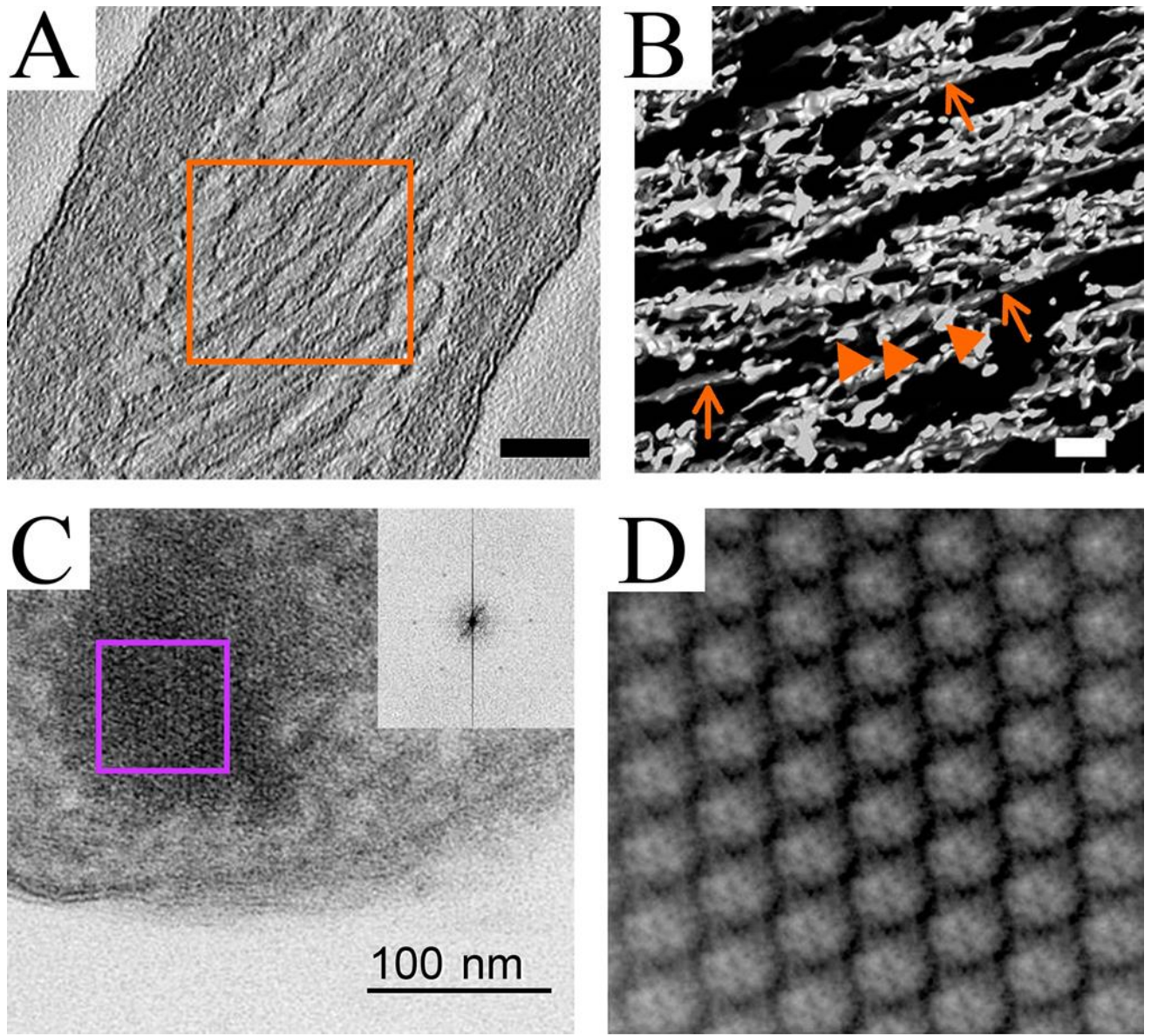

Fig. 6. Variety of condensed nucleoid structures in dormant *E. coli* cells. (A) Central section through tomogram of *E.coli* cell with quasi-liquid crystalline structure. Bar – 100 nm. (B) 3D representation of liquid crystals, marked on (A) with an orange frame. Arrows are pointing to the DNA, arrowheads – to the Dps. (C) TEM of the quasi- nanocrystalline structure inside *E.coli* cell (semi-thick section); insert – Fourier transform; (D) filtered co-crystal;

Discussion

Dormant cells are formed by bacteria depleted from nutrients for longer periods of time (Minsky et al., 2002; Sinitsyn et al., 2017; Wolf et al., 1999). Most cells (up to 99.98%) in long-term starving populations undergo autolysis. The remaining cells develop into anabiotic dormant forms, which differ significantly in structural organization from vegetative cells. So far, the mechanism of condensation of nucleoid into above mentioned structures is unclear.

Using thin sections electron microscopy and semi-thin sections electron tomography, we demonstrated the variety of structures in dormant *E.coli* cells, namely: quasi - nanocrystalline, quasi-liquid crystalline (or spore - like)and the novel, described here for the first time, folded nucleosome-like structures. The first two types of structure were described in previous investigations (Wolf et al., 1999; Minsky et al., 2002; Frenkiel-Krispin & Minsky, 2006; Sinitsyn et al., 2017; Loiko et al., 2017; Krupyanskii et al., 2018). Quasi-liquid crystalline structure has been found in the cytoplasm of mutants lacking a *dps* gene (Minsky et al., 2002). This structure was similar to cholesteric liquid-crystalline phase of DNA. We have found quasi-liquid crystalline (or spore - like) structures in bacterial cells populations with a different environmental conditions and Dps content small but not equal zero (Table), as in the case of(Minsky et al., 2002). Structure represented in Fig. 4 A and Fig. 6 A, B is close to cholesteric liquid-crystalline phase of DNA, found by (Minsky et al.,2002). Rod - like DNA inside quasi - liquid crystalline structure (Fig. 6 A, B) (confirmed by the presence of phosphorus on the elemental analysis) may interact with individual Dps molecules (confirmed by the presence of sulfur in the elemental analysis); each Dps is a spherical dodecamer with a diameter of ~10 nm, which coincides with the particle size on our tomogram (arrowheads on Fig. 6B). Absolutely different structure for cells displayed on the Fig. 4 (B, D). Such type of condensed DNA structures is closer to the structure of nucleoids in spores (Loiko et al., 2017). In most cells, these DNA structures are surrounded by a dense material, which makes them even more similar to bacterial spores (Loiko et al., 2017). DNA molecules in these structures are randomly distributed. Core of the cell where DNA molecules are situated contains small amount of DPS. The number of quasi-liquid crystalline (or spore - like) structure is higher in populations grown in less abundant environments, and decreases with the rise of Dps content (Table).

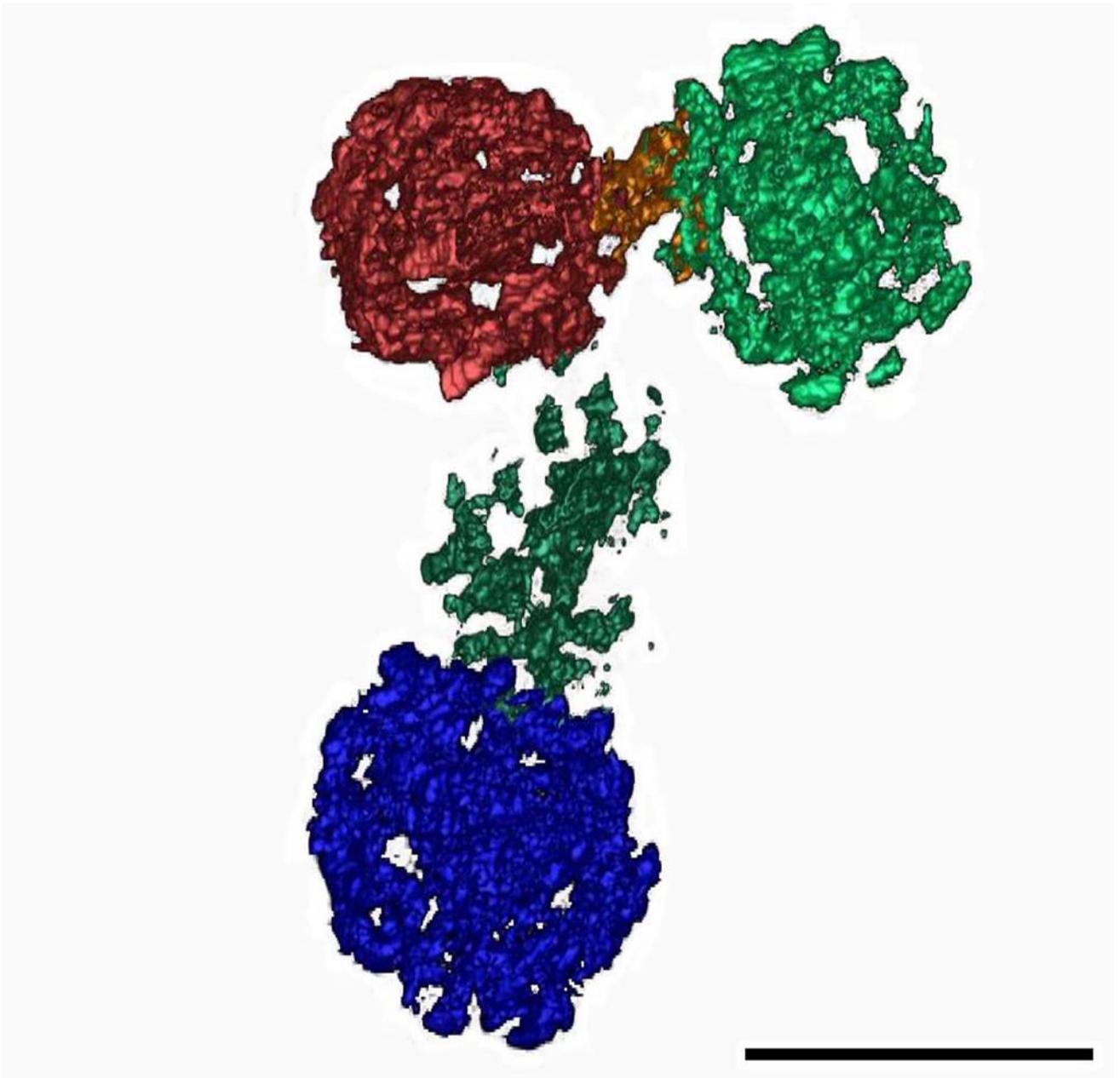

Fig. 7. Three-dimensional image of associated regions on a Fig. 5B (of the three to the left). Blue, red and green are beads, or spherical associates of DPS molecules with an average diameter of about 30 nm.

Of particular interest is the third type of structure since it was described here for the first time: the folded nucleosome-like. *Folded nucleosome-like structures* are more often presented in cells that grow on a synthetic media (Table).
Tomographic study (Fig. 5A,B) clearly demonstrated that these structures are not toroids which are the intermediate form in the formation of the quasi - nanocrystalline structure (Frenkiel-Krispin & Minsky, 2006), but represent much smaller and nearly spherical formations. It was suggested that in bacteria cells spherical aggregates of Dps molecules arrange the DPS beads (Fig.7) through which the DNA (string) passes (beads on the string (Fig.8A). For protection of external stress factors, the beads are placed quite tightly on the DNA (string). Similar to eukaryotic cells, where the nucleosomes are folded to form fiber loops, which then form the chromatin, beads on the string may folded into a globule-like structure. Possible schematic presentation of folded nucleosome - like structure, suggested here may be seen on Fig 8 (B, C,D).

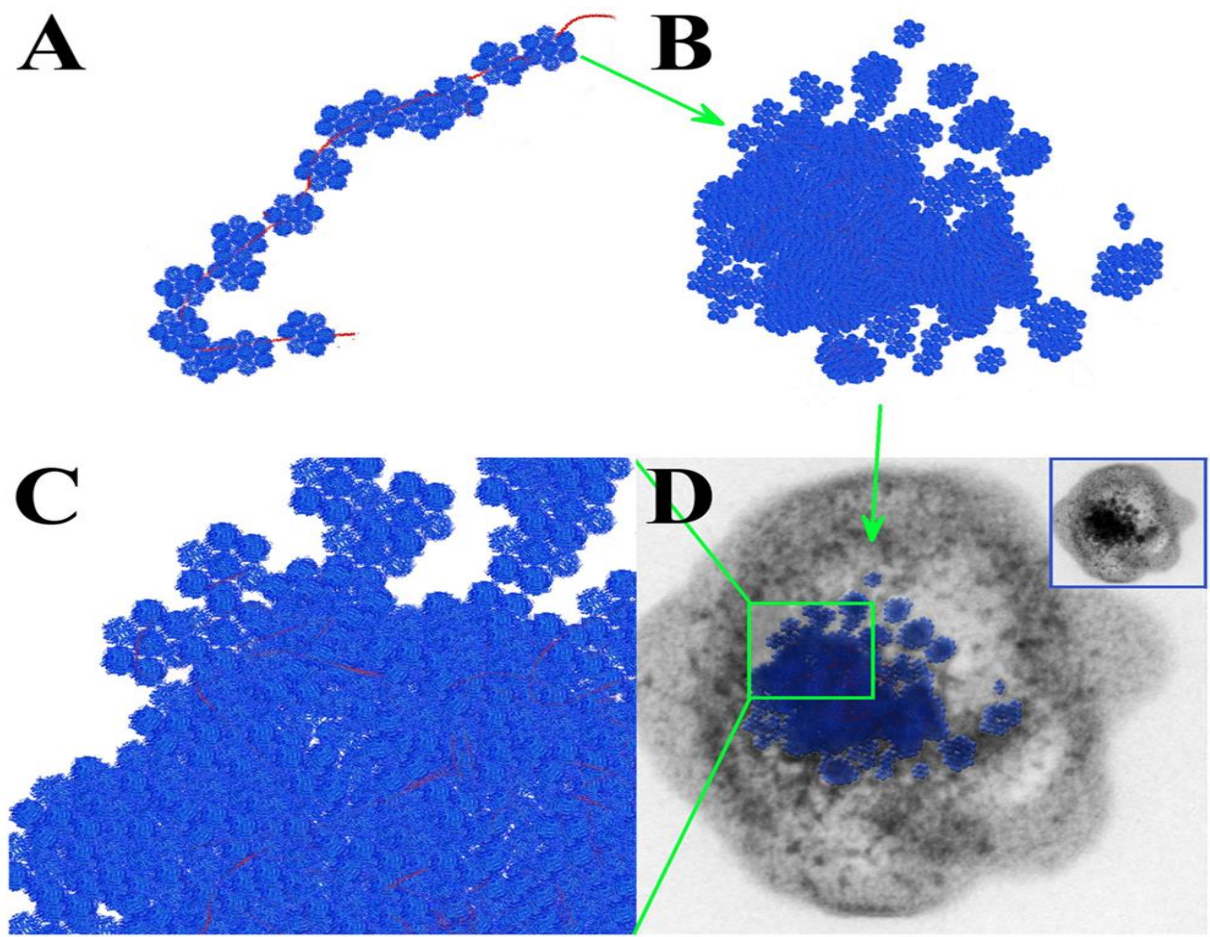

Fig.8. Possible schematic representation of folded nucleosome - like structure (extension of Fig.7). (A) Beads on the string, where string is a DNA and beads are as in Fig.7. For better protection of DNA from the external environment, the chain (beads on the string) is folded into a globule-like structure (B). Part of the globule with additionally adherent DPS molecules is shown in ( C). A full image of the nucleosome-like globule superimposed with the cell is shown in (D). The inset shows a cell without a schematic globule.

Results observed here shed a new light both on the phenomenon of nucleoid condensation in prokaryotic cells and on the general problem of developing a response to stress and another unfavorable conditions by microorganisms. The fact that cells bearing abovementioned forms of nucleoid condensation were found in populations aged around 7 months proves that they are not transitional intermediate forms, but that their formation is programmed in a development cycle to implement different survival strategies.

Finally, it was find out that there is no single mechanism for nucleoid condensation in the population of a dormant cell; diversity in their number shape and packing have been seen. According to the recognized concept of a bacterial population as a multicellular organism (Shapiro & Dworkin, 1997), its heterogeneity allows to respond flexibly to the environmental changes and to survive in stressful situations. That is the reason why we observed at least three types of nucleoid condensation in dormant *E. coli* cells. Heterogeneity of dormant cells increases the ability of the whole population to survive under various stress conditions.

For better understanding of the nucleoid condensation mechanism, it is necessary to study the reverse transition of the nucleoid in bacteria from the dormant to the functional state.

Experimental Procedures

The subjects of research were gram-negative bacteria *Escherichia coli* K12 and *E.coli* Top10 (wild type), and the Dps over-producer strains *E.coli T*op10/pBAD-DPS and *E. coli* BL21-Gold(DE3)/pET-DPS.

Plasmid constructing

The DNA fragment encoding DPS was obtained by PCR amplification of *E.coli* K12 MG1655 DNA using the following oligonucleotides:
dps-nde 5'- GATATGAACATATGAGTACCGCTAAATTAG
dps-hind 5'- TATAAGCTTATTCGATGTTAGACTCGATAAAC

The DNA fragment was introduced into the plasmid pET-min at the restriction sites NdeI and HindIII. The obtained plasmid pET-DPS contains the DNA region encoding the DPS protein under the control of the T7 promoter. *E.coli* strain BL21-Gold (DE3) was transformed with the pET-DPS plasmid.

For the pBAD-DPS, the recipient vector pBAD/Myc-His A was used. The promoter was removed from the plasmid by cleaving the restriction sites BamHI and HindIII. The remote promoter region was obtained by PCR amplification and seamlessly combined with the DPS coding region by PCR. The removed plasmid region was amplified on the pBAD/Myc-His A template using oligonucleotides PB-F and PB-dpsR. The DNA fragment encoding DPS was obtained by PCR amplification of E. coli K12 MG1655 DNA using PB-dpsF and dps-hind oligonucleotides. DNA fragments were purified by preparative agarose gel electrophoresis, combined and amplified using oligonucleotides PB-F and PB-dpsR. The resulting DNA fragment was inserted into the plasmid pBAD/Myc-His A using the restriction sites BamHI and HindIII. The result is a plasmid pBAD-DPS containing the DNA region encoding the DPS protein, under the control of the promoter of the arabinose operon. Strain *E.coli* Top10 was transformed with plasmid pBAD-DPS.

Electrophoresis

The bacterial cells were suspended in water and disintegrated in an ultrasonic sonicator. The lysate was centrifuged at 13000g for 10 min. The supernatant was analyzed by SDS-PAAG.

Dormant forms of E. coli Top10 were obtained by growing bacteria for 24 h at 33°C under shaking (140 rpm) in 250-mL flasks with 50 mL of LB medium. Then cells were stored at 21ºC for 7 months.

Dormant forms of E. coli K12 were obtained by cultivation for 14 days at 28°C without agitation in 750-mL flasks with 300 mL of the modified M9 medium with decreased ammonium nitrogen content. The samples were concentrated 20-fold in the same medium and stored in plastic test tubes for 7 months at 21°C.

Dormant forms of E. coli Top10/pBAD-DPS were obtained in two ways:
1) Bacteria were grown for 24 h at 28°C under shaking (140 rpm) in 250-mL flasks with 50 mL of modified M9B medium. Dps overproduction was induced by 1 g/l of arabinose in the linear growth phase. After that cells were stored for 7 months at 28ºC.
2) Bacteria were grown for 24 h at 28°C under shaking (140 rpm) in 250-mL flasks with 50 mL of LB medium. Dps overproduction was induced by 1 g/l of arabinose in the linear growth phase. Then cells were stored at 21 ºC for 7 months.

Dormant forms of E. coli BL21-Gold(DE3)/pET-DPS were obtained by growing bacteria for 24 h at 28°C under shaking (140 rpm) in 250-mL flasks with 50 mL of modified M9 medium with decreased ammonium nitrogen content. Dps overproduction was induced by 10 мМ of lactose in the linear growth phase. The samples were stored for 7 months at 21ºC.

Light microscopy was carried out using a Zetopan light microscope (Reichert, Austria) under phase contrast.

Sample preparation for electron microscopy. Cells were fixed with 2% glutaraldehyde for 5 hours and postfixed with 0,5 % paraformaldehyde; washed with a 0.1 M cacodylate buffer (pH=7.4), contrasted with a 1% $OsO_4$ solution in a cacodylate buffer (pH=7.4), dehydrated in an increasing series of ethanol solutions, followed by dehydration with acetone, impregnated and embedded in Epon-812 (in accordance with manufactures instructions). Ultrathin sections (100 – 200 nm thick) were cut with a diamond knife (Diatome) on an ultramicrotome Ultracut-UCT (Leica Microsystems), transferred to copper 200 mesh grids, covered with formvar (SPI, USA) and contrasted with lead citrate, according to the Reynolds established procedure (Reynolds, 1963). For analytical electron microscopy(EM) study, contrasting was in some cases omitted.

Transmission electron microscopy. Ultra-thin sections were examined in transmission electron microscopes JEM1011 and JEM-2100 (Jeol, Japan) with accelerating voltages 80 kV and 200kV, respectively, and magnification of x13000-21000. The images were recorded with Ultrascan 1000XP and ES500W CCD cameras (Gatan, USA). Tomograms were obtained from semi-thick (300-400 nm) sections using *Jeol Tomography* software (Jeol, Japan). The tilting angle of a goniometer ranges from -60 ° to +60 ° (with a permanent step of 1 degree). Series of images were aligned by Gatan Digital Micrograph (Gatan, USA) and then recovered with the back-projection algorithm in IMOD 4.9. Three-dimensional sub-tomograms were visualized in the UCSF Chimera package (Pettersen et al., 2004).

Analytical electron microscopy was carried on an analytical transmission electron microscope JEM-2100 (Jeol, Japan), equipped with a bright field detector for scanning transmission electron microscopy (SPEM) (Jeol, Japan), a High Angular Angle Dark Field detector (HAADF) (Gatan, USA), a X-Max 80 $mm^2$ Silicon Drift Detector (Oxford Instruments, Great Britain) and a GIF Quantum ER energy filter (Gatan, USA). STEM and TEM modes were used. The STEM probe size was 15 nm. EDX spectra and element analyses were performed in the INCA program (Oxford Instruments, Great Britain).

The X-ray diffraction measurements were performed using synchrotron radiation at the ID23-1 beamline of the ESRF (Grenoble, France) at a wavelength λ = 1.6799 Å, the slit width was 10 μm, the exposure time was 5 s per frame, the PILATUS 6M two-dimensional detector was placed at a distance of 95 cm behind the sample in the incident-beam line. The measurements were performed at 100 K.

Acknowledgements

Authors would like to thank Lisa Trifonova for proofreading the manuscript. Research was performed within frameworks of the state tasks for ICP RAS 0082-2014-0001 (state registration #AAAA-A17-117040610310-6) and 0104-2018-0029. Analytical electron microscopy and dual-axis tomography was performed at the User Facility Center "Electron microscopy in life sciences" of Moscow State University. We also thank Sankt-Petersburg University Resource Center for Molecular and Cellular Technologies and Alexandra Ivanova for help with the use of the JEOL2100 electron microscope. The authors thank the European Synchrotron Radiation Facility for providing the possibility of conducting experiments.